\documentclass[fleqn,usenatbib,useAMS]{mnras}

\usepackage{graphicx,epsfig}
\usepackage{url}

\usepackage{amsmath}
\usepackage{amssymb}

\usepackage{threeparttable}

\usepackage{ifsym}

\usepackage{xcolor}

\newcommand{\kms}{\,km\,s$^{-1}$}
\newcommand{\msun}{M$_{\odot}$}

\title[The Chemically Peculiar Star $\theta^1$~Ori~F]{\'Echelle Spectroscopy of the Chemically Peculiar Star $\theta^1$~Ori~F \thanks{Based on
observations obtained at the Observatorio Astron\'omico Nacional
at San Pedro M\'artir, Baja California, M\'exico, operated by the
Instituto de Astronom\'{\i}a, Universidad Nacional Aut\'onoma de
M\'exico.}}

\author[Costero et al.]{Rafael Costero$^1$\thanks{In active retirement; E-mail: costero@astro.unam.mx }, Christine Allen$^1$,  Alex Ruelas-Mayorga$^1$, Leonardo S\'anchez$^1$  
 \newauthor
   Julio Ram\'irez V\'elez$^2$, Juan Echevarr\'ia$^1$ and Gustavo C. Melgoza$^2$, \\
    $^1$Instituto de Astronom{\'\i}a, Universidad Nacional Aut\'onoma de M\'exico,
Apartado Postal 70-264, 04510 M\'exico, D.F.,  M\'exico \\
    $^2$Instituto de Astronom{\'\i}a, Universidad Nacional Aut\'onoma de M\'exico,
Apartado Postal 877, 22830, Ensenada, Baja California, M\'exico \\
}

\date{Accepted XXX. Received YYY; in original form ZZZ}
\pubyear{2020}

\begin{document}

\label{firstpage}
\pagerange{\pageref{firstpage}--\pageref{lastpage}}

\maketitle


\begin{abstract}
 
We analyse \'Echelle spectra of $\theta^1$~Ori~F obtained by us on six nights unevenly distributed along six years; we identify several hundred spectral lines and measure, for the first time, the star's heliocentric radial velocity. We also collect and discuss previously published photometry of $\theta^1$~Ori~F.  
  
We find that $\theta^1$~Ori~F is a Chemically Peculiar (CP) star with overabundant silicon and phosphorus, and possibly other elements as well. From the singly ionised Fe, Cr and Ti lines we estimate its spectral type to be between B7 and B8. The radial velocity of $\theta^1$~Ori~F is  possibly  marginally variable, with an average of $24 \pm 4.2$\kms\,(standard deviation), in good agreement with the mean radial velocity of the Orion Nebula Cluster members, and about 5\kms\ smaller than the average of the other Trapezium components. We cast doubt on the coeval nature of this star relative to the other Trapezium components, and present arguments that almost certainly exclude its membership to the Orion Trapezium.  $\theta^1$~Ori~F turns out to be enigmatic in several respects, and is probably an important link for understanding the evolutionary stage at which the CP phenomenon sets on.

\end{abstract}


\begin{keywords}
stars: individual (Orion Trapezium Cluster) -- stars: individual ($\theta^1$~Ori~F) -- stars: chemically peculiar --  techniques: spectroscopic.
\end{keywords}


\maketitle

\section{Introduction}
\label{sec:intro}

$\theta^1$~Ori~F = Brun 603 = Parenago 1892 is a $V$=10.15 star 4.5 arcsec away from Component C, the brightest member of the Orion Trapezium and the main ionising source in the Orion Nebula. Very little is known about this intermediate magnitude star. 
Recently its membership to the Trapezium has been questioned; hence our increased interest to learn about the poorly known kinematic and physical properties of this star in order to assess its disputed physical relation to the Orion Trapezium. 

The fifth and sixth stars in the Orion Trapezium,  $\theta^1$~Ori E and F \citep[ADS 4186 E and F, according to the nomenclature adopted in the Double Star Catalogue by][]{ait32}, were usually omitted or poorly considered in most early photographic and photometric studies of the Orion Nebula Cluster (ONC).  This occurred because of their proximity to much more luminous members of the crowded stellar group, and their location in the brightest part of the emission nebula. They were also frequently excluded from recent, higher spatial resolution surveys of the ONC, because of their excessive brightness as compared with the vast majority of the cluster members.
Consequently, there was a conspicuous gap in the physical information about these intermediate mass members of the very young stellar cluster, an important deficiency given their possible proximity to the zero age main sequence.

This lack of information is particularly remarkable in the case of $\theta^1$~Ori F. The star is not included in the radial velocity survey of the ONC members by \citet{sicilia05}, nor in more general surveys like the {\scshape SDSS}, or Gaia. Only preliminary values of its radial velocity were given in \citet{oea13} and \citet{cos19}. In the present work our definitive value is provided. 
Furthermore, its spectral type, B8, was only known from a short comment at the end of a paper by \citet{her50} on the spectral classification of variable stars in the ONC. In the same paragraph Component E was also given a classification, but it was later recognized that the assigned composite type was wrong \citep{hergrif06}. These uncertain classifications were registered, and hence perpetuated, by \citet{par54} in his photographic survey of the ONC, where Component F was assigned a photovisual magnitude of 11.0.

Ever since their discovery in 1826 -- attributed to W.\,Struve by \cite{Web59} -- the brightness of components E and F, estimated mostly by eye, has been suspected to vary.  In a short communication, \cite{gle80a} proposed that the visibility of these relatively faint components was highly dependent on atmospheric conditions (seeing and transparency).  Later that year, \cite{gle80b} encouraged amateur observers to monitor both stars for their variability, albeit warning them that previously reported variations in brightness were based on their relative brightness.
\cite{bru35}, in his careful cartographic work of the Orion Nebula region, placed special attention on the variability of stars, but neither $\theta^1$~Ori~F nor E were included in his list of variable ONC members. However, the early reports of visually detected relative variation and further photometric evidence of possible variability of component E in the visual range \citep[e.g.][]{wal77, fei78}, led to the permanence of both E and F in the revised version of the Catalogue of Suspected Variable Stars under numbers 2291 and 2296, respectively \citep{kuk81}.
However, in the only paper we know of in which PSF photometry of the bright Orion Trapezium stars is reported, \cite{Wol94} found no significant variations for Component F,  whereas for Component E this author reported ``definite night to night variations of several tenths of a magnitude" relative to Component D.
Surprisingly, in their recent infrared survey devoted to find variability in the ONC members using the {\it Spitzer} telescope for multiepoch observations in the 3.6 and $4.5\,\micron$  bands, \cite{mor11} did not report any measurements for $\theta^1$~Ori~F. The absence of Component F in this survey is strange, since it contains brighter stars and, of course, many fainter ones.

$\theta^1$~Ori F is the only B-type star in the ONC (out of 12) that was not detected as an X-ray source in the Chandra Orion Ultradeep Project \citep{ste05, get05, prei05}. The star has not been detected in radio--frequencies either  \citep[e.g. see][]{gar87, felli93, zap04, koun14}.

No circumstellar matter or disk has been registered around  $\theta^1$~Ori F; an upper limit of $0.12 M_{jup}$ was established by \cite{man14} for the mass of a disk around this star.
However, at about $2\arcsec$ from it and almost exactly in the opposite direction from where $\theta^1$~Ori C is located, lies the interesting double proplyd LV~1 \citep[168-326\,NW and 168-326\,SE in][]{OdW94}. These three objects --C, F and LV~1-- are in a probably fortuitous alignment \citep [see figures in][]{gra02, Wu13, Wu18}.

During the science verification tests of the GRAVITY VLTI instrument, \cite{mcc16} found a stellar object very close (about 8 mas) to $\theta^1$~Ori F.
 However, this provisional result was recently disproved by the \cite{GRAVITY18}. These authors set a 3-sigma upper limit of 1.5 \msun\ for any interferometric companion separated more than 1.7\,AU (4 mas) from the primary star. 
 These findings encouraged us to analyse and measure \'Echelle optical spectra we had previously obtained for this star. In Section 2 of this paper we report the results of our analysis; in particular, we provide the radial velocities measured for the star, upper limits to its projected velocity of rotation and magnetic mean modulus, and other data drawn from those spectra. In an attempt to better assess the physical properties of $\theta^1$~Ori F and its membership to the Orion Trapezium, we also collect and briefly discuss, in Section 3, the published CCD photometry of this mostly ignored star. In Section 4 we argue about the $\theta^1$~Ori F evolutionary status and its membership to the Orion Trapezium in particular, and to the ONC in general. Our conclusions are given in Section 5.


\section{Spectral Analysis}
\label{sec:spectroscopy}


\subsection{Observations}
\label{sec:spec1}

$\theta^1$~Ori~F was observed during six nights between 2007 and 2013 using the \'Echelle spectrograph, with resolving power R\,$\sim$\,15000, at the f/7.5 Cassegrain focus of the 2.1\,m telescope of the Observatorio Astron\'omico Nacional at San Pedro M\'artir (OAN-SPM), B.C., M\'exico. Three different detectors were used: during the first three nights, the ``SITe3''  Photometrics CCD (1024$\times$1024, 24\,\micron\,pixels) was employed; in the fourth night a Thomson CCD with 2048$\times$2048, 14\,\micron\,pixels was used; and in the last two nights, the e2v-4240 ``Marconi\,1'' CCD, with 2048$\times$2048 pixels, 13.5\,\micron\,in size, was utilised. In all the observations the 300\,l/mm \'Echellette grating was used to cover the spectral range spanning from 3800\,\AA\ to 6900\,\AA. The slit was always set at position angle $30^\circ$ to minimize the dispersed light contamination from Component C.

Exposure times varied according to the quality of the sky and the availability of observing time (borrowed from other research programs), as detailed in the log of observations listed in Table \ref{tab:LogSpec}.
Each spectrum, or group of consecutive spectra, was flanked by wavelength calibration provided by a Th\,Ar lamp, with the exception of those obtained on the first and the third nights (2007 Nov 23 and 2008 Jan 19), for which only one wavelength calibration was acquired at the end of the exposure.
Standard {\scshape IRAF}
\footnote{IRAF is distributed by the National Optical Astronomy Observatories, which are operated by the Association of Universities for Research in Astronomy, Inc., under cooperative agreement with the National Science Foundation.}
procedures were used to reduce the data. No flat-fielding, flux calibration or dispersed light subtraction were applied.


\begin{table}
\caption{Log of Spectroscopic Observations}
\label{tab:LogSpec}
\begin{center}
\setlength{\tabcolsep}{1.45\tabcolsep}
\begin{tabular}{cccc}
\hline
Date & HJD & Exp$^{*}$ \\
(dd-mm-yyyy UT) & (2 450 000 +)	&   (s) \\
\hline
23-11-2007 & 4427.9208 & 1600         \\
18-01-2008 & 4483.7534 & 900$\times$3 \\
19-01-2008 & 4484.7360 & 900          \\
17-01-2011 & 5578.7403 & 900          \\
21-01-2013 & 6313.7585 & 900$\times$3 \\
28-11-2013 & 6624.8897 & 1200$\times$3 \\

\hline
\end{tabular}
\end{center}
{* Exposure time}
\end{table}

\subsection {Spectral Description and Classification}
\label{sec:spec2}


 The spectrum of $\theta^1$~Ori~F shows relatively shallow hydrogen absorption lines (contaminated by their counterparts arising from both the nebula and Component C) and numerous very narrow metallic lines. With the help of the B-type synthetic stellar spectra, available from C. Gummersbach \& A. Kaufer\footnote{https://www.lsw.uni-heidelberg.de/projects/hot-stars/websynspec.php)} and of the {\scshape NIST-ASD} compilation \citep{kra18}, we were able to identify several hundreds of them in the co-added spectrum. This spectrum was produced by averaging the four longest exposed spectra listed in Table \ref{tab:LogSpec}, after normalising and shifting them to rest velocity. Obvious nebular lines were excised. Most of the co-added spectrum is shown in Appendix  \ref{sec:echellespectra}.  Some important features are briefly described below.

In the observed spectral range, the Mg\,${\rm II}$ $\lambda$\,4481\,\AA\ line is the strongest metallic line, at least doubling the equivalent widths of the next five strongest ones: Si\,${\rm II}$ $\lambda\lambda$\,4128, 4131, 5056, 6347 and 6371\,\AA. Other Mg\,${\rm II}$ lines are also clearly detected. 
The Fe\,${\rm II}$ lines, some of them very strong, are by far the most numerous. Several lines of Fe\,${\rm III}$ are also detected, albeit very weakly. 
The expected He\,${\rm I}$ lines, based on the B8 classification of this star by \cite{her50}, are either absent or completely filled-in by their nebular counterparts. Traces of the corresponding relatively wide lines due to contamination from Component C are sometimes discernible. 
The C\,${\rm II}$ $\lambda$4267\,\AA~line, which has a behaviour similar to that of the He\,${\rm I}$ lines in normal B0\,-\,A0 stars, is faintly detected in our co-added spectrum of $\theta^1$~Ori~F.

\begin{figure*}
\centering
\includegraphics[width=17cm]{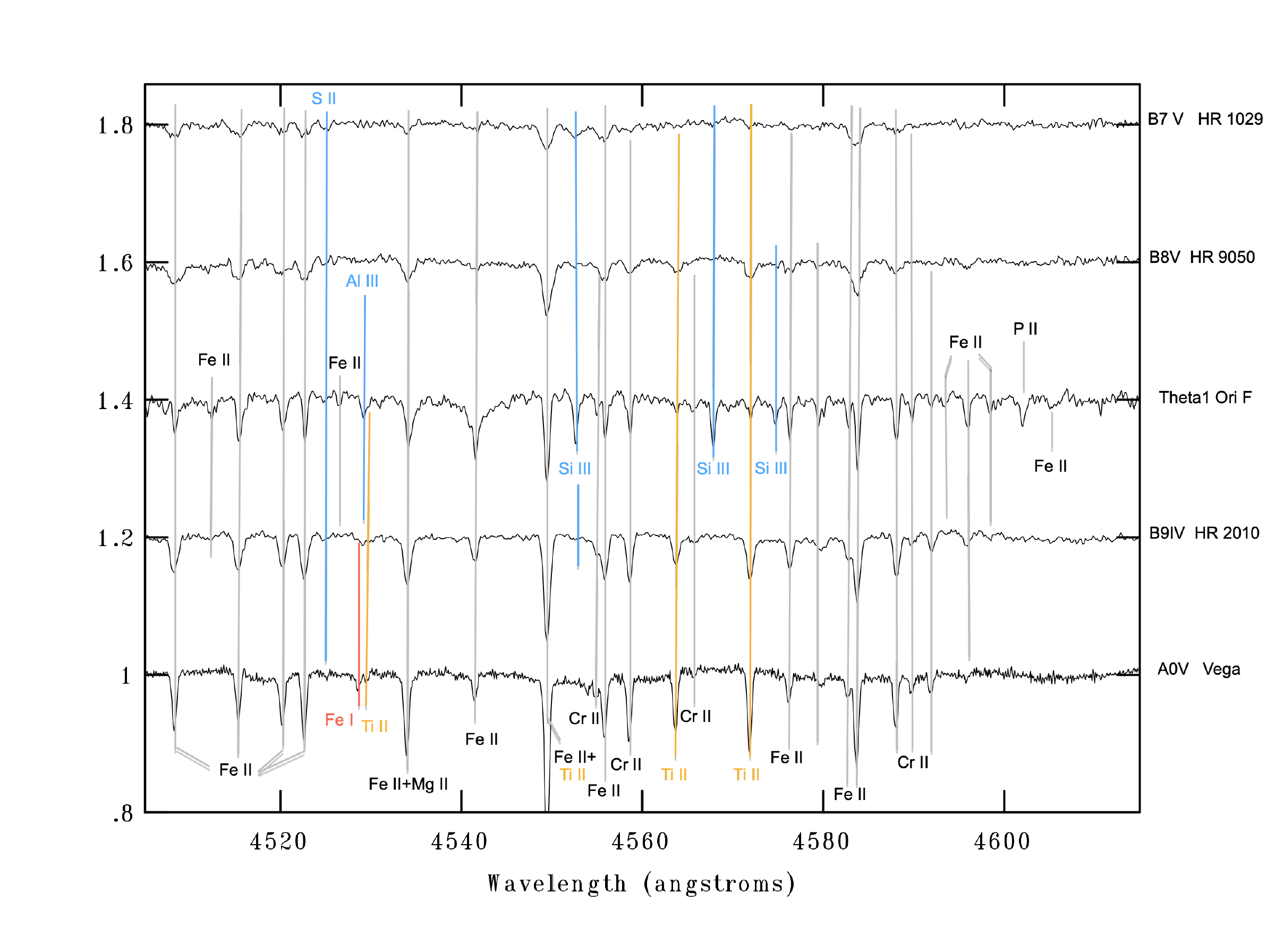}
\caption{Normalised co-added spectrum of $\theta^1$~Ori~F compared with narrow-line B7--A0 spectral standards, in the 4505-4615\,\AA\ interval. The spectra are shifted by 0.2 units in the vertical axis in order to facilitate their visualisation, and in wavelength to bring them all to rest radial velocity. Lines that are prominent in $\theta^1$~Ori~F, but very weak or absent in the standard stars, notably the Si\,${\rm III}$ triplet and the P\,${\rm II}$ line, $\lambda$\,4602\,\AA, are identified in the spectrum of Component F. The presence of Cr\,${\rm II}$ lines and the weakness or absence of the Ti\,${\rm II}$ lines in $\theta^1$~Ori~F, suggests that its spectral type is around B7.5.}
\label{fig:FvsStands4505-4615A}
\end{figure*}

\begin{figure*}
\centering
\includegraphics[width=17cm]{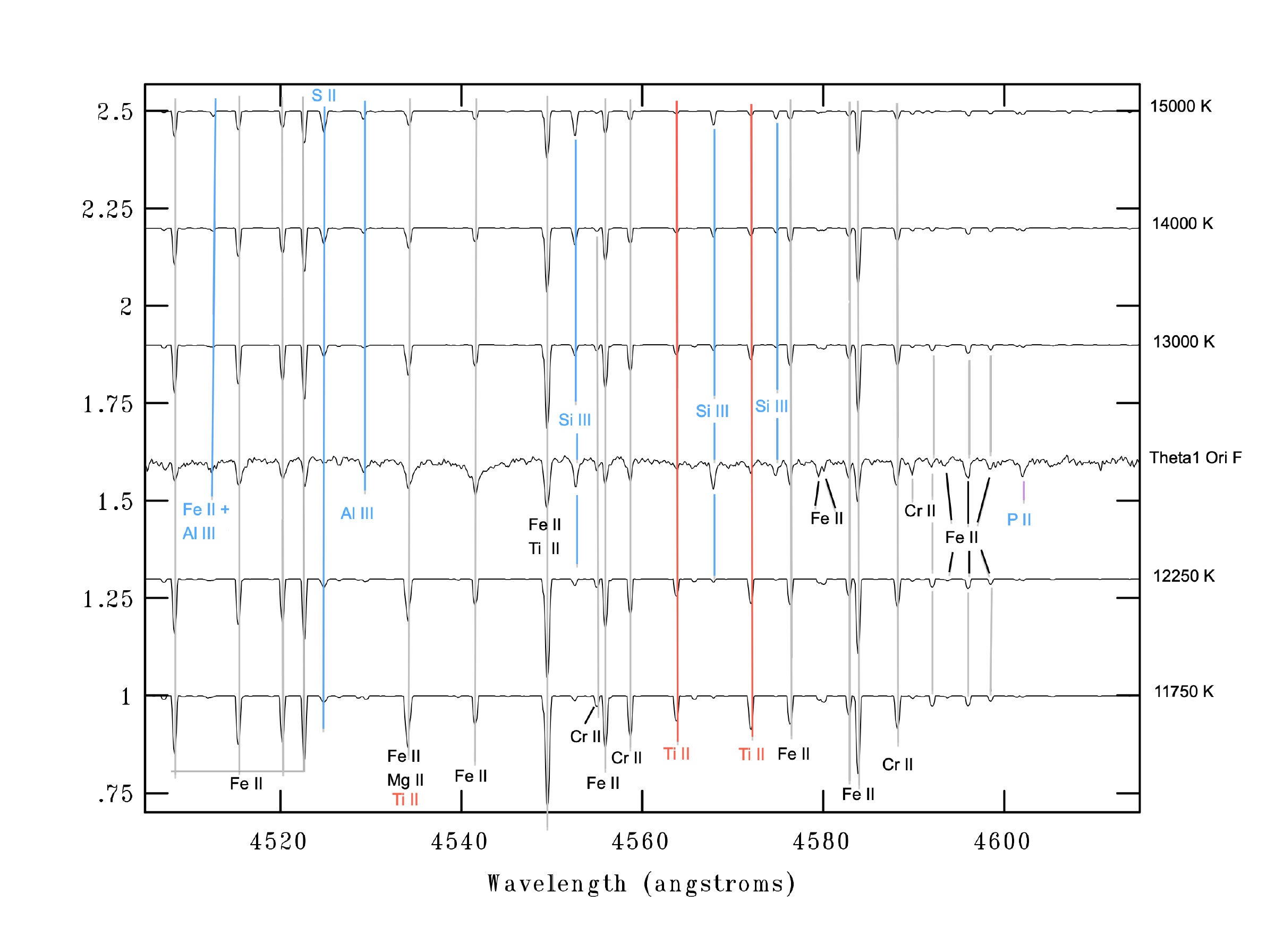}
\caption{Normalised co-added spectrum of $\theta^1$~Ori~F compared with synthetic spectra with $log~g = 4.0$ and 0.3\,~dex silicon overabundance in the 11750\,K -- 15000\,K temperatures range. Note that the Si\,${\rm III}$ triplet is stronger in $\theta^1$~Ori~F than in any of the synthetic spectra. The wings at both sides of Fe\,${\rm II}$ $\lambda$\,4542\,\AA\ in Component F are due to contamination from the He~II line in the spectrum of the nearby, much brighter and hotter Component C.}
\label{fig:FvsSint0_3Si4505-4615}
\end{figure*}

Outstanding is the relatively large strength of the Si\,${\rm II}$ lines originated from the higher excitation levels (e.g. $\lambda\lambda$\,4201, 4377, 4673, 5467, 5800, 5868, 6240, 6661, 6672\,\AA.) Indeed, the presence of Si\,${\rm II}$ $\lambda$\,4201\,\AA\ in the spectrum of a star was used by \cite{sea64} to classify it as an Ap\,Si star.
Furthermore, as shown in Figs. \ref{fig:FvsStands4505-4615A} and \ref{fig:FvsSint0_3Si4505-4615}, the Si\,${\rm III}$ triplet $\lambda\lambda$\,4553, 4568 and 4575\,\AA, strong in normal early-B-type stars and nearly absent in types B7 and later, is clearly present in $\theta^1$~Ori~F. 
These abnormalities (including the apparently weak He\,${\rm I}$ and C\,${\rm  II}$ lines) are typical of chemically peculiar (CP) stars.

Also peculiar are the strong P\,${\rm II}$ lines in $\theta^1$~Ori~F (e.g.  $\lambda\lambda$\,4602, 6024, 6034 and 6043\,\AA).  Probably also enhanced are the Al\,${\rm III}$ lines at $\lambda\lambda$\,4512.6 and 4529.2\,\AA\ , and those of Sr\,${\rm II}$ at $\lambda\lambda$\,4077.7 and 4215.5\,\AA. There is no trace of O\,${\rm II}$ absorption lines, many of which could be filled-in by nebular recombination emission \citep[see][]{esteban+04}. 

 Several Cr\,${\rm II}$ lines, which appear around B6 type stars, are clearly present in $\theta^1$~Ori~F. Additionally, the Ti\,${\rm II}$ lines, which are very weak in B7 stars and become prominent in A0 stars (e.g. $\lambda\lambda$\,4300, 4444, 4564 and 4572\,\AA) are very weak or barely discernible from the noise in the co-added spectrum of $\theta^1$~Ori~F, as shown in Fig. \ref{fig:FvsStands4505-4615A}.


Very few neutral metallic lines are present in $\theta^1$~Ori~F. Only the O\,${\rm I}$ lines around $\lambda$\,6157\,\AA\ are prominent; the strongest Fe\,${\rm I}$ lines are barely detected in the co-added spectrum and there is no trace of the Mg\,${\rm I}$ triplet around $\lambda$\,5175\,\AA, nor of Ca\,${\rm I}$ $\lambda$\,4227\,\AA.  
There are several broad and shallow lines, most of them around the position of the He\,${\rm I}$ and He\,${\rm II}$ lines (see, e.g., $\lambda$4542\,\AA\  in Figs. \ref{fig:FvsStands4505-4615A} and \ref{fig:FvsSint0_3Si4505-4615}) caused by light contamination from Component C. Some other lines (like the one around 6280\,\AA ) are of telluric origin, smeared in the co-addition process. There is a noticeably broad absorption feature for which we do not have a convincing explanation: the one around the Na\,${\rm I}$ D interstellar doublet (see Fig. A3). This feature is not present in Component C, observed on two occasions just after Component F, and hence probably it is not a consequence of the lack of flat field correction; yet, we cannot rule out the possibility of its being an artifact of the reduction process, especially when normalising the spectra. We can also speculate that this enigmatic line, if real, could stem from the wide and strong Na\,${\rm I}$ doublet produced in an as yet undetected pre-main sequence companion to $\theta^1$~Ori~F, some 3 to 4 $V$ magnitudes fainter than F, with a spectral type between K5 and M0, and situated in a low-inclination, relatively close orbit. Such a hypothetical star would not be detected, neither by the \citet{GRAVITY18} nor by us. However, no other strong line typical of late-type stars was found in our co-added spectrum. A similar binary system (CP primary and pre-main sequence secondary) has been observed in AO Vel \citep{Gonzal06}.


The observed spectral peculiarities of $\theta^1$~Ori~F preclude the precise classification of the star in the Morgan \& Keenan system. However, based on the weak but clear presence of the Cr\,${\rm II}$ lines and the barely detectable Ti\,${\rm II}$ ones, we conclude that it is a CP B7.5\,p star, with an overabundance at least of Si and P. This classification is in good agreement with the B8 spectral classification assigned by \citet{her50} to this star from a single, medium-dispersion photographic spectrum.


\subsection{Projected Stellar Rotation and Interstellar Lines}
\label{sec:spec3}

Limited by the spectral resolution of our instrument (about 0.45\,\AA) and by the modest signal-to-noise ratio of our spectra, we can estimate only an upper limit for the projected equatorial velocity of rotation of $\theta^1$~Ori~F,  v\,sin\,i $\lesssim$ 20\,\kms. This is supported by the slightly narrower width of its spectral lines as compared with those in Vega, observed by us with the same instrumental setup, and for which \cite{roy02} measured v\,sin\,i = 24\,\kms.

As a CP B7.5\,p Si star, $\theta^1$~Ori~F is expected to possess a strong magnetic field; however, its detection requires a much higher spectral resolution than the one at present available to us. The relatively young age of $\theta^1$~Ori~F highlights the importance of much higher resolution spectral studies of this peculiar star.

One spectrum of the nearby, much brighter $\theta^1$~Ori~C was also obtained immediately after that of $\theta^1$~Ori~F in each of the last two observing nights. After reduction and normalisation, these spectra were compared with those of Component F to evaluate the consequences of the light contamination of the fainter star by the brighter one. Additionally, we compared the depths and shapes of the interstellar Na\,${\rm I}$ D doublet in both stars, and noticed they are almost identical (aside from the very wide component mentioned above), indicating that the neutral gas between us and both stars is quantitatively and dynamically very similar.


\subsection{Synthetic Spectra and Temperature Estimate}
\label{sec:spec4}

In an attempt to reproduce the absorption lines of $\theta^1$~Ori~F we used the code {\sc Cossam}\footnote{COdice per la Sintesi Spettrale nelle Atmosfere Magnetiche} to create synthetic stellar spectra \citep[e.g.][]{sti12}. This code is especially designed to analyse the stellar magnetised atmospheres by solving the polarised radiative transfer equations, i.e. by calculating the four disk integrated Stokes parameters (I,Q,U,V). However, since no sizeable magnetic field can be detected using medium spectral resolution, we adopted a non-magnetic atmospheric model in {\sc Cossam}.

For the synthesis of the spectra, we selected different atmospheric models from the Atlas9 grid by \citet{cas2004}, with $T_{\rm eff} = [11750\ - 16000]$\,K, log $g$ = 4.0 and a micro turbulence value $V_{\rm turb}$\,= 2\,\kms.
The line formation parameters were obtained from the VALD database \citep{ryab15}.
Solar abundances reported in \citet{asp09} were adopted, except for that of silicon, for which we adopted 0.3, 0.6 and 1.2\,dex above solar abundance.

No synthetic spectrum is able to reproduce the observed intensities of all the silicon lines. Apparently, the higher excitation and ionisation Si lines arise mainly from spots or bands in the star's surface, where this and other elements are enhanced by magnetic diffusion, as has been suggested by \cite{michaud+81}. These spots and bands have been detected in some CP stars through Doppler tomography and maximum entropy techniques by several observers \citep[e.g. see][and references therein]{Hatzes93}, who also were unable to fit simultaneously the Si~II and Si~III lines with only one effective temperature. The intensities of the lower Si\,${\rm II}$ excitation lines relative to the neighbouring Fe\,${\rm II}$ lines are better adjusted with an 0.3\,dex Si overabundance, which we adopt herein.

The effective temperature of a star may also be estimated by measuring the ratios of equivalent widths of suitable pairs of lines arising from the same ion, but from atomic levels with significantly different excitation energies \citep[e.g. see][and references therein]{catal02}. These ratios can be compared with those measured in synthetic spectra calculated for different temperatures.  
However, this method is not appropriate in the case of $\theta^1$~Ori~F because lines arising from low-excitation atomic levels are unexpectedly weaker in $\theta^1$~Ori~F than lines arising from higher excitation ones. Hence, their ratios yield temperatures higher than $15000$\,K. Also, the ratio of Si\,${\rm II}$ to Si\,${\rm III}$ lines yields temperatures even higher than those obtained from Fe\,${\rm II}$ line ratios. These discrepancies are typical of CP stars and have been explained by vertical abundance gradients in their atmospheres \citep [see][and references therein]{ryab14}. 

In brief, there is no single temperature that can characterise the surface of $\theta^1$~Ori~F. 
The chemical peculiarities of the star prevent us from using mixed element line intensities to reliably estimate its temperature,  However, if the Cr/Ti abundance ratio is normal, then the star cannot be cooler than about 12000\,K, or hotter than 14000\,K. Hence, from its spectroscopic data alone, the spectral type of  $\theta^1$~Ori~F can be best described as CP B7.5\,p\,Si, with an uncertainty of 1 in its spectral sub-type.


\subsection{Radial Velocity}
\label{sec:spec5}

The radial velocity of $\theta^1$~Ori~F was not known previously.  Its value is of great relevance to discern if this star is a member of the Orion Trapezium or not.  The original purpose of this work was to determine it. To this end, we cross-correlated its spectrum with a suitable synthetic one.
The {\scshape IRAF} task {\em fxcor} was used to perform the cross-correlation on each of 18 adjacent \'Echelle spectral orders of the six spectra listed in Table \ref{tab:LogSpec}. Obvious nebular lines were excised from the object's spectra.
The wavelength interval for each of the 18 \'Echelle orders was selected so as to include the whole corresponding free spectral range, with no overlap between adjacent intervals. The total spectral interval spans from 4120\,\AA\ to 6470\,\AA. The Balmer $H\gamma$ and $H\beta$ lines, the [O\,${\rm I}$] $\lambda$\,5577~\AA\ telluric emission line and the Na\,${\rm I}$ $D_1$ and $D_2$ interstellar lines were excluded from the correlation.
We selected as synthetic spectrum the one calculated for $T_{\rm eff}=12750$\,K and 0.3\,dex silicon overabundance, because it yields the largest correlation height in the 4500--4625\,\AA\ interval (where many lines of species in different ionisation states are present) when the object's spectrum is cross-correlated with all the synthetic spectra we created with the adopted silicon overabundance.

In the second column of Table \ref{tab:RadVel} we list the heliocentric radial velocities of $\theta^1$~Ori~F, measured on the Julian dates listed in the first column. Each of these velocities is the average of 18 radial velocities measured in the same number of \'Echelle orders in each spectrum, weighted by their corresponding TDR value \citep[a confidence coefficient defined by][]{tonrydavis79}, which is calculated by the {\em fxcor} task together with the other correlation parameters.
In the third column, the standard deviation from the mean of the 18 individual velocities is listed.


\begin{table}
\caption{Heliocentric Radial Velocity Measures for $\theta^1$~Ori~F}
\label{tab:RadVel}
\begin{center}
\setlength{\tabcolsep}{1.45\tabcolsep}
\begin{tabular}{cccc}
\hline
HJD &  Vrad  &  StDv  &  Vcorr\\
(2 450 000 +) & ($\rm km\,s^{-1}$) & ($\rm km\,s^{-1}$) & ($\rm km\,s^{-1}$) \\
\hline
4427.9208   & 26.9 & 2.8 & 29.7 \\
4483.7534   & 18.9 & 2.7 & 22.7 \\
4484.7360   & 23.2 & 2.5 & 26.3 \\
5578.7403   & 26.2 & 4.2 & 22.2 \\
6313.7585   & 21.8 & 2.7 & 17.5 \\
6624.8897   & 28.4 & 2.7 & 25.7 \\
\hline
Average     & 24.3 & 2.9 & 24.0 \\
Stand. Dev. &  3.6 & --- &  4.2 \\
\hline
\end{tabular}
\end{center}
\end{table}


The last column of Table \ref{tab:RadVel} lists the same radial velocity, but corrected by a zero-point shift that was estimated as follows.
In each spectrum, the average heliocentric radial velocity of strong, non saturated, emission lines from the Orion Nebula was calculated. The lines used for this purpose were H$\beta$, He\,{\rm I} $\lambda\lambda$\,4471 and 5875\,\AA, and O\,{\rm III} $\lambda$\,4959\,\AA. The zero-point correction is the difference between the average radial velocity derived from these lines and that interpolated by us at the position of $\theta^1$~Ori~F from those obtained by \cite{cast88} over most of the Orion Nebula using the [O\,{\rm III}] $\lambda$ 5007\,\AA\ line. 
The latter velocity was estimated to be 18.5\,\kms, but instead we adopted 19.5\,\kms\, because the wavelength used by \citet{cast88} for the [O\,{\rm III}] line was 0.017\,\AA\ shorter than that presently listed in the {\em NIST} compilation, our main source of wavelength information for almost all the stellar lines (see below). Inspection of Table \ref{tab:RadVel} shows that these zero-point corrections, usually smaller than 4\,\kms, caused a very small decrease in the average heliocentric velocity, but they somewhat increased its standard deviation.

In order to check the reliability of the synthetic spectrum used for the cross-correlation process and the zero point corrections adopted, we also evaluated the radial velocity of $\theta^1$~Ori~F by directly measuring the wavelength shift of about 120 lines, deemed adequate for this purpose, chosen from many absorption lines previously identified. The resulting velocities are very similar to those obtained with the cross-correlation method; their average and standard deviation are 24.9\,$\pm$\,3.6\,\kms\,  and 24.7\,$\pm$\,5.0\,\kms\, before and after applying the zero-point corrections described above. Within errors, these values are equal to those listed at the bottom of Table \ref{tab:RadVel}, obtained by the cross-correlation method.

Although these results are subject to improvement, they point to a possibly variable radial velocity; alternatively, they may be caused by line profile variations due to the rotational modulation of a spotty surface structure.


\section{Published photometry}
\label{sec:photometry}

To our knowledge, there is no published aperture photometry of $\theta^1$~Ori F; this is understandable, since the star is too close to a much brighter one and inside the brightest part of the Orion Nebula. It is surprising that, in spite of the advantages of modern detectors, almost no CCD image photometric data in the visual range are available for this star.  Improved infrared array technology, with even higher spatial resolution, has favoured its observation in the near-infrared, specifically in the H and K bands.  In Table \ref{tab:PubPhot} we list all the original CCD photometric measurements for $\theta^1$~Ori F found in the literature. We then briefly discuss them and draw some conclusions regarding a possible photometric variability and a peculiar spectral distribution.


\begin{table*}
\caption{Published photometry of $\theta^1$~Ori~F.}
\label{tab:PubPhot}
\begin{center}
\setlength{\tabcolsep}{2.4\tabcolsep}
\begin{tabular}{cccccccccc}
\hline
 U  &  B  &  V  &   $I_C$  &  J  & H &  K  &  L  &  References  \\
\hline
 --- & --- & 10.18 & 9.2 & --- & --- & 9.0 & --- & 1,2,3 \\
9.928* & 10.075* & 10.122 & 9.652* & ---  & --- &  8.86 & --- & 4,4,4,4,5 \\
--- & --- & --- & --- & --- & 9.05 & 8.99 & --- & 6,6 \\
--- & --- & --- & --- & --- & 9.309 & 9.142 & --- &  7,7 \\
--- & --- & --- & --- & 8.68 & 8.50 & 8.38 & 8.81 &  8,8,8,8 \\
--- & --- & --- & --- & --- & --- & --- & 8.935 &  9 \\

\hline
\end{tabular}

{1) Prosser et al. (1994), 2) McCullough et al. (1995) 3) McCaughrean \& Stauffer (1994), 4) Da Rio et al. (2009) 5) Simon et al. (1999), 6) Petr et al. (1998), 7) Hillenbrand \& Carpenter (2000), 8) Muench et al. (2002), 9) Lada et al. (2004).}\\
{* The U, B and I magnitudes by Da Rio et al. are not in the Johnson or Cousins systems, and are not colour corrected.}

\end{center}
\end{table*}

\subsection{Analysis of the Photometric Data}
\label{sec:photanal}

$\theta^1$~Ori E and F were listed by \cite{pro94} in Table 7 of their HST photometric survey of the ONC. This frequently ignored table contains the $V$ magnitude of nine non-saturated ``bright'' stars in the Trapezium region, measured on an image exposed for only one second. There, Component F was assigned $V = 10.18$, a value close to that found later by \cite{der09}, who performed nearly simultaneous ground based CCD photometry of the ONC with $UBVI$ filters.  Note that these $U$, $B$ and $I$ magnitudes do not correspond to any defined photometric system, and no colour correction was applied.  
It is remarkable that the colour excess $E\,(B-V) = 0.06\pm0.04$ --derived from this photometry and adopting the intrinsic colour index for normal B8V stars, $(B-V)_0 = -0.109$ (\citealt{pec13}, see also the webpage\footnote{http://www.pas.rochester.edu/\string~emamajek/ 
\\ EEM\_dwarf\_UBVIJHK\_colors\_Teff.txt})-- is much smaller than the average colour excess measured for the three brightest members of the Orion Trapezium, $E~(B-V) \sim 0.32$ \citep{lee68, car88}. Indeed, if this average colour excess is assumed to apply also to $\theta^1$~Ori F, then the observed $(B-V)=-0.045$\,mag for this star by \citet{der09} would imply an intrinsic $(B-V)_0=-0.365\,$mag, bluer than that of a black body at infinite temperature \citep{arp61}. 
So, either this star is subject to a substantially lower interstellar extinction than that of the other Trapezium stars or the observations at the extreme blue are considerably contaminated by the close Component C and the surrounding nebular emission.
As we describe in Section \ref{sec:spec2}, $\theta^1$~Ori F is a chemically peculiar (CP) star. These stars frequently have $(B-V)$ colours bluer than their normal spectral-type counterparts, \citep[e.g. see][]{Leckrone73}, though the anomaly is no larger than 0.2 \,mag \citep[as in 41\,Tau, HR\,1732 and 108\,Aqr in][]{sea64}. Hence, the CP character of $\theta^1$~Ori F could only partly explain its excessively blue $(B-V)$ colour inferred from the photometry published by \citet{der09}. 

The large dispersion in the infrared photometric data published for $\theta^1$~Ori F could be explained by the star being variable in brightness. Yet, as mentioned in Section \ref{sec:intro}, \cite{Wol94} found no appreciable variation of the brightness of Component F (relative to Component D) in his study of the bright Trapezium stars. During 25 nights distributed over two years, \cite{Wol94} obtained more than 100 images in each of the six filters ($u,v,b,y,R_C,I_C$) used. Unfortunately, only the plots of the period-folded light curves of Component B --one for each filter-- were published. From these curves we estimate the photometric errors to be clearly less than 0.1\,mag, particularly in filters $y$ and $I_C$. Hence, we may safely conclude that \cite{Wol94} found no variations larger than 0.1\,mag in $\theta^1$~Ori F in the observed spectral range.

In contrast, the five $K$ magnitudes listed in Table \ref{tab:PubPhot} range from 8.38 to 9.14, a difference much larger than the reported observational errors. Note that only the brightest value --- given by \cite{mue02} --- differs significantly (by 0.64~mag) from the average $<K>=9.00\pm0.14$~mag of the other four.  However, the $H-K$ colour indices --- from the three groups that obtained nearly simultaneously both $H$ and $K$ magnitudes --- are very similar to each other, averaging $0.116\pm 0.038$.  This colour index is consistent with that of a B7 star with 1.6 or 1.8\,mag of visual interstellar extinction, under either a normal ($R=3.1$) or an Orion Trapezium ($R=5.3$) extinction law \citep{car89}.

The fairly large putative variability inferred from some of the photometric observations of $\theta^1$~Ori F is not expected for a main-sequence, late B-type star, even a CP one. Considering the observational challenges for this star, it is plausible that substantial errors are the cause of at least part of the differences. But it is also possible for a yet undetected, close companion to $\theta^1$~Ori F (or a nearby background star) ---a late K or early M pre-main-sequence star--- undergoing flare-type variability, to be the culprit of the suspected variability. Flare stars are abundant among the low--mass members of star forming regions, such as the ONC and the OB associations around it.


\begin{figure}
\centering

\includegraphics[width=12cm]{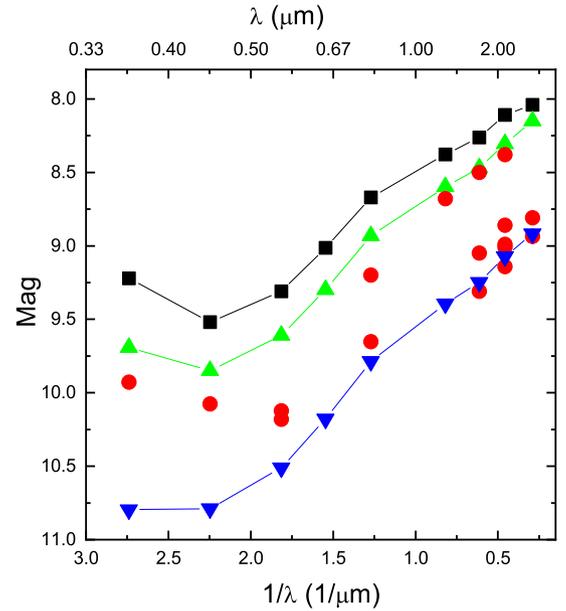}
\caption{Expected spectral distribution of main sequence B6 (black squares), B8 (green triangles) and B9 (blue inverted triangles) stars, located at the distance of the Orion Trapezium. The observed magnitudes for $\theta^1$~Ori~F, listed in Table \ref{tab:PubPhot}, are represented by red circles. We have adopted the distance to the Orion Trapezium as 400 pc, an $A_V=1.8$ mag, and $R_V=5.3$. }
\label{fig:Int_Mag_B6B8B9_Obs_vs_1_lambda}
\end{figure}


Both the abnormally blue spectral distribution and the large dispersion of the near infrared magnitudes of $\theta^1$~Ori~F are illustrated in  in Fig. \ref{fig:Int_Mag_B6B8B9_Obs_vs_1_lambda}. There we plot the observed magnitudes for $\theta^1$~Ori~F (listed in Table \ref{tab:PubPhot}), together with the spectral distribution of late B--type main sequence stars (B6V, B8V and B9V) placed at a distance of 400\,pc (a reasonable compromise between the distance given by \citealt{men07} and \citealt{koun17}) and applying the reddening corrections given by \citet{car89} with  $A_V = 1.8$\,mag and $R_V = 5.36$ (the average values found for the Trapezium stars by the latter authors). The intrinsic magnitudes and colours for the hypothetical stars were taken from \citet{pec13}.  
The observed magnitudes lie between those expected for normal B8V and B9V stars under the conditions described above. But the increasingly large brightness excess observed in the $V$, $B$ and $U$ filters, relative to the $I$ magnitude, compared with the expected spectral distribution for a normal B9 main sequence star, suggests the possibility that this CP star has an abnormally large V, B and U excesses and/or that the observations are contaminated by the nearby Trapezium Component C; this is consistent with the fact that the bluer the measured magnitude, the larger is the brightness excess.  

More intriguing are the subluminous values thst are equal, within errors, for the $L$ magnitude of $\theta^1$~Ori~F obtained by \citet{mue02} and \citet{lada04} . As shown in Fig. \ref{fig:Int_Mag_B6B8B9_Obs_vs_1_lambda} this $L$ magnitude is consistent with the $H$ and $K$ magnitudes measured by other authors (that of a B9V star), but is about 1.5\,mag fainter than that expected from the $J$, $H$ and $K$ values obtained nearly simultaneously by \citet{mue02}. 
A possible explanation for the $L$-band subluminosity could be related to erroneous subtraction of the bright and patchy background due to the emission band around 3.3\,\micron, originated in small dust grains (PAHs) excited by the strong UV radiation of the nearby OB stars.


\section{Membership of the F component to the Orion Trapezium}
\label{sec:memb}

Traditionally, Component F has been considered as part of the Orion Trapezium. It is listed as such in catalogues of visual binary and multiple stars, such as the ADS, the IDS and the WDS. However, the large transverse velocity with respect to several of the Trapezium components found from both historical measures and using diffracto-astrometry measures on Hubble images \citep{oli12} rendered doubtful its physical association to this multiple system \citep{oea13}.

\citet{all74, all04} devised several tests to establish the physical membership of moving components to their trapezia. Using only data on the relative planar positions over a time base of more than 150 years, they applied these tests to the Orion Trapezium components A to F.  These tests were: (i) a statistical test, involving the density of field stars in the vicinity of the system; (ii) a proper motion test, comparing the direction and magnitude of the transverse velocity of the component with the proper motion of a companion; and (iii) a luminosity function-proper motion test, designed to exclude foreground stars.  These tests are described in detail in \cite{all04}.  We re-applied them specifically to Component F.  Given the close proximity of F and C, the first test indicated that physical membership is probable. Furthermore, the stars in the Orion Trapezium are known to be in front of the very dense molecular cloud OMC-1 \citep[e.g. see][]{Ode01}.  This renders highly improbable the presence of a background star so close to Component C.  The second test was found to be inconclusive,  because of the persisting uncertainties in the proper motion of Components A, B, C, D. 
For the third test, \cite{all04} obtained that the expected number of  foreground stars present among their entire sample of  40 trapezium systems (which included the Orion Trapezium) was 0.4.  This test indicated that the probability of F being a foreground object was small, but not zero.  The original test did no take into account the then unknown population of foreground stars discovered by  \citet{abo12}, which clearly  increases the likelihood of F being a foreground star.
Based solely on the kinematic information on Component F, it is thus not possible to unambiguously confirm its physical membership to the Orion Trapezium. The rather large value for the transverse velocity of Star F with respect to A (12.4\,\kms), to B (7.5\,\kms) and to C (11.2\,\kms) found by \cite{oli12} would argue in favour of its being a foreground star. The luminosity function-proper motion test would marginally allow F to be a foreground star.  Indeed, \cite{oea13} concluded that F was a foreground star, probably related to the stellar population found by \cite{abo12} and located in front of the ONC.

In view of the conflicting information regarding the membership of Component F to the Orion Trapezium, we decided to consider further tests.
The determination of the radial velocity of F, presented in this paper, shows its value to be  24.0\,\kms, in good agreement with the peak in the radial velocity distribution of the ONC members found by \citet{sicilia05} and \citet{fuerenz08} (25 and 26\,\kms, respectively) and somewhat smaller than the average systemic velocity of the four brightest Orion Trapezium components, about 29\,\kms \citep{all17}. Again, solely from this information, we cannot conclusively exclude or include its membership to the Orion Trapezium.

A concordant parallax of Star F with that of the other members of the Trapezium would favour its membership. Future Gaia parallaxes may be useful to settle this question.  However, at present the Gaia DR2 parallaxes for stars A, C, D and F (which imply distances of 421, 369, 472 an 404\,pc, respectively) are very uncertain, due to the fact that all Trapezium stars (except possibly Component F) are close multiples, and hence not reliably treated by Gaia DR2.

In this context it is pertinent to note that, as shown in Section \ref{sec:spectroscopy}, $\theta^1$~Ori~F is a chemically peculiar (CP) star with overabundant silicon and phosphorus, and possibly other elements as well. 
This behaviour points toward Component F being an older, evolved star. Even if it is still a zero-age main sequence star and its mass is less than 4\,\msun, it should be older than about $10^6$ years according to \citet{sie00}\footnote{http://www.astro.ulb.ac.be/\string~siess/pmwiki/\\pmwiki.php?n=WWWTools.Isochrones} and \citet{scha92}. This is clearly incompatible with the very young dynamical age of about 40000 yr found both for the Orion Trapezium as a whole \citep{all17}, and for its B component \citep{all15}.
If  Component F should turn out to be indeed a member and, hence, coeval to the other Trapezium members, then the very young age implied by the dynamical models would make it an extraordinary object, important for understanding the stage at which the CP phenomenon sets on.  The quadruple system in AO Vel \citep{Gonzal06}, and $\sigma$ Ori E -- the helium-enhanced, rapidly rotating CP magnetic star also located in a young trapezium system -- may be additional relevant examples of very young CP stars. 

It is interesting to note that \citet{bec17} found evidence for three distinct star populations within the ONC, and that \citet{krou18} developed a model for a staggered star formation within nearly coeval groups like the ONC, where successive generations of stars may form in bursts, each lasting about $10^5$ years and separated by intervals of about a million years. This scenario may be relevant to the understanding of peculiarities such as those shown by $\theta^1$~Ori~F.


\section{Conclusions}
\label{sec:concl}

From the \'Echelle spectra, acquired over a six-year lapse, we obtained for the first time the radial velocity of $\theta^1$~Ori~F and found that it is a chemically peculiar (CP) star. The mean value of its radial velocity, $24 \pm 4.3$\,\kms\,(standard deviation), is slightly smaller than the average estimated for the other Orion Trapezium components. This standard deviation is barely larger than the expected one sigma level (3\,\kms) and it could be partially caused by line-profile variations similar to those observed in other CP stars. 
Hence, we conclude that our results show no spectroscopic binarity in $\theta^1$~Ori~F. This result complements the upper limit (at a 3-sigma level) set by the \citet{GRAVITY18}, which precludes the existence of a stellar companion to $\theta^1$~Ori~F separated by more than 1.7\,AU from the primary star, and with $M>1.5$\,\msun; specifically, we can establish that there is no binary companion to $\theta^1$~Ori~F with $M>0.8$\,\msun\ closer than 1.7\,AU, assuming the best conditions (i.e., an edge-on orbit). 

Because of its CP character, the star cannot be neatly classified in the Morgan--Keenan system, though it may be reasonably described as a CP\,B7-8 Si star, with an overabundance at least of silicon and phosphorus.
Much better spectral resolution is needed to measure its magnetic field intensity and its projected rotational velocity, for which we could determine only an upper limit of 20\,\kms\ .

The membership of $\theta^1$~Ori~F to the Trapezium has been recently cast in doubt on the basis of its transverse motion relative to other Trapezium components \citep{oli12,oea13}. 
In favour of non-membership we can now add further arguments: its chemical peculiarities are generally found only in evolved stars, and point to Component F being much older than the other Trapezium stars.  Indeed, the dynamical lifetime of the Orion Trapezium as a stellar group was recently found to be at most 40000 years \citep{all17}.
In addition, near infrared photometry indicates that it is near the main sequence and of type between B8 and B9; hence, its mass should be between 2.8 and 5.3\,\msun\,.  But then its evolutionary state is hard to reconcile with that of other Trapezium members of similar masses
--- namely, the two practically identical components of the spectroscopic binary $\theta^1$~Ori~E, each with a mass of $2.80\,\pm\,0.05$\,\msun\ \citep{mor12}, and the secondary component of the eclipsing binary $\theta^1$~Ori~A = V1016 Ori, with a mass of about 3.9\,\msun\ \citep{vitriplach01} or 3.0\,\msun\, \citep{stick00,cos19} --- all of which are definitely pre-main sequence stars. 
Indeed,  pre-main sequence evolutionary models \citep{scha92,sie00}  indicate that it takes a 3.0\,\msun\,star approximately $2.4 \times 10^6$ years to reach the main sequence, whereas a 5.0\,\msun\,star reaches the beginning of its hydrogen burning phase in about $8 \times 10^5$ years. These times are at least an order of magnitude larger than the estimated age of the Orion Trapezium.

Weighing all the evidence, we conclude that $\theta^1$~Ori~F is not a physical member of the Orion Trapezium, but a star that could have been formed in another episode within the same nebula, in a manner similar to that envisioned by \citet{krou18}, who propose that star formation proceeds in staggered bursts within ONC-like groups. If this is the case, Component F could indeed be part of the older populations of the ONC noted by \citet{abo12} and  \citet{bec17}, still within the Orion Nebula but not in the Trapezium.


\section*{Acknowledgements}

We are very grateful to our referee, Dr Tatiana Ryabchikova, for many constructive comments and suggestions, which resulted in an improved paper. Thanks are due to DGAPA (Direcci\'on General de Asuntos del Personal Acad\'emico at UNAM) for partial financial support under projects PAPIIT IN-102517, IN-103320, IN-102617 and IN-103120.
We also thank Juan Carlos Yustis for his help with some of the figures presented in this paper.

This research has made use of the SIMBAD database, operated at CDS, Strasbourg, France.


\section*{Data availability}
The data underlying this article can be shared on request to the corresponding author.


%

\begin{thebibliography}{}

\bibitem[Allen et al.(1974)]{all74} Allen, C., Poveda, A. \& Worley, C.\,E. 1974, RevMexAA, 1, 101

\bibitem[Allen et al.(2004)]{all04} Allen, C., Poveda, A. \& Hern\'andez-Alc\'antara, A., 2004, RevMexAA(SC), 21, 195

\bibitem[Allen et al.(2015)]{all15} Allen, C., Costero, A. \& Hern\'andez, M., 2015, AJ, 150, 167

\bibitem[Allen et al.(2017)]{all17} Allen, C., Costero, R., Ruelas-Mayorga, A. \& S\'anchez, L.\,J., 2017, MNRAS, 466, 4937

\bibitem[Aitken \& Doolite(1932)]{ait32} Aitken, R.\,G. \& Doolittle, E. 1932, {\it New General Catalogue of Double Stars within 120$^{\circ}$  of the North Pole}, Carnegie Institution of Washington

\bibitem[Alves \& Bouy(2012)]{abo12} Alves, J. \& Bouy, H. 2012, A\&A, 547, A97

\bibitem[Arp(1961)]{arp61} Arp\,H., 1961, ApJ, 133, 874

\bibitem[Asplund et al.(2009)]{asp09} Asplund, M., Grevesse, N., Sauval, A.\,J., et al. 2009, ARA\&A, 47, 481

\bibitem[Brun(1935)]{bru35} Brun, A. 1935, Pub. Obs. Lyon, 1, 12

\bibitem[Beccari et al.(2017)]{bec17} Beccari, G., Petr-Gotzens, M.\,G., Boffin, H.\,M.\,J. et al. 2017, A\&A, 604, A22

\bibitem[Cardelli \& Klayton(1988)]{car88} Cardelli, J.\,A. and Clayton, G.\,C. 1988, AJ, 95, 516

\bibitem[Cardelli et al.(1989)]{car89} Cardelli, J.\,A., Clayton, G.\,C. and Mathis, J.\,S., 1989, ApJ, 345, 245

\bibitem[Castaneda(1988)]{cast88} Castaneda, H.\,O. 1988, ApJS, 67, 93

\bibitem[Castelli \& Kurucz(2004)]{cas2004} Castelli, F., and Kurucz R.\,L. 2004, ArXiv e-prints[arXiv:astro-ph/0405087]

\bibitem[Catalano et al.(2002)]{catal02} Catalano, S., Biazzo, K., Frasca A., and Marilli, E. 2002, A\&A, 394, 1009 

\bibitem[Costero(2019)]{cos19} Costero, R., 2019, ArXiv e-prints[arXiv:astro-ph/1906.11956]

\bibitem[Da Rio et al.(2009)]{der09} Da Rio, N., Robberto, M., Soderblom, D,\,R. et al. 2009, ApJS, 183, 261

\bibitem[Esteban et al.(2004)]{esteban+04} Esteban, C., Peimbert, M., Garcia-Rojas, J. et al. 2004, MNRAS, 355, 229

\bibitem[Feibelman \& Gull(1978)]{fei78} Feibelman, W.\,A. \& Gull, T.\,R. 1978, PASP, 90, 762

\bibitem[Felli et al.(1993)]{felli93} Felli, M., Taylor, G.\,B., Catarzi, M. et al. 1993, A\&AS, 101, 127

\bibitem[F{\H{u}}r{\'e}sz et al.(2008)]{fuerenz08} F{\H{u}}r{\'e}sz, G., Hartmann, L.\,W., Megeath, T. et al. 2008, ApJ, 676, 1109

\bibitem[Garay, Moran \& Reid(1987)]{gar87} Garay, G., Moran, J.\,M. \& Reid, M.\,J. 1987, ApJ, 314, 535

\bibitem[Getman et al.(2005)]{get05} Getman K.\,V., Flaccomio, E., Broos, P.\,S. et al. 2005, ApJS, 160, 319

\bibitem[Gledhill(1880a)]{gle80a} Gledhill, J. 1880a, Astronomical register, 18, 64

\bibitem[Gledhill(1880b)]{gle80b} Gledhill, J. 1880b, Astronomical register, 18, 313

\bibitem[Gonz\'alez et al.(2006)]{Gonzal06} Gonz\'alez, J.\,F., Hubrig, S., Nesvacil, N., North P., 2006, A\&A, 449, 327

\bibitem[Graham et al.(2002)]{gra02} Graham, M.\,F., Meaburn, J., Garrington, S.\,T., et al., 2002, ApJ, 570, 222

\bibitem[GRAVITY collaboration et al.(2018)] {GRAVITY18} Gravity Collaboration, Karl, M., Pfuhl, O. et al. 2018, A\&A, 620, A116

\bibitem[Hatzes(1993)]{Hatzes93} Hatzes, A.\,P. 1993, {\it ASP Conference Series}, 44, 258

\bibitem[Herbig(1950)]{her50} Herbig, G.\,H. 1950, ApJ, 111, 15

\bibitem[Herbig \& Griffin(2006)]{hergrif06} Herbig, G.\,H. \& Griffin, R.F. 2006, AJ, 132, 1763

\bibitem[Hillenbrand \& Carpenter(2000)]{hil00} Hillenbrand, L.\,A. \& Carpenter, J. 2000, ApJ, 540, 236

\bibitem[Kounkel et al.(2014)]{koun14} Kounkel, M., Hartmann, L., Loinard, L., et al., 2014, ApJ, 790, 49

\bibitem[Kounkel et al.(2017)]{koun17} Kounkel, M., Hartmann, L., Loinard, L., et al., 2017, ApJ, 834, 142

\bibitem[Kramida et al.(2018)]{kra18} Kramida, A., Ralchenko, Yu., Reader, J. and NIST ASD Team, 2018. {\it NIST Atomic Spectra Database (version 5.5.6),} [Online]. Available: https://physics.nist.gov/asd . National Institute of Standards and Technology, Gaithersburg, MD.

\bibitem[Kroupa et al.(2018)]{krou18} Kroupa, P., Je{\v{r}}{\'a}bkov{\'a}, T., Dinnbier, F., et al. 2018, A\&A, 612, A74

\bibitem[Kukarkin et al.(1981)]{kuk81} Kukarkin, B.~V., Kholopov, P.~N., Artiukhina, N.~M., et al. 1981, {\it Catalogue of suspected variable stars}, Moscow, Acad. of Sciences USSR Sternberg

\bibitem[Lada et al.(2004)]{lada04} Lada, C.~J., Muench, A.~A., Lada, E.~A., et al. 2004, AJ, 128, 1254

\bibitem[Lee(1968)]{lee68} Lee, T.\,A. 1968, ApJ, 152, 913

\bibitem[Leckrone(1973)]{Leckrone73} Leckrone, D.\,S. 1973, ApJ, 185, 577

\bibitem[Mann et al.(2014)] {man14} Mann, R.~K., Di Francesco, J., Johnstone, D., et al. 2014, ApJ, 784, 82


\bibitem[McCaughrean \& Stauffer(1994)]{mcc94} McCaughrean, M.\,J. \& Stauffer, J.\,R. 1994, AJ, 108, 1382

\bibitem[McCaughrean(2016)]{mcc16} McCaughrean, M.\,J. 2016,\\ http://www.eso.org/public/images/eso1601f/

\bibitem[McCullough et al.(1995)] {mcc95} McCullough, P.~R., Fugate, R.~Q., Christou, J.~C., et al. 1995, ApJ, 438, 394

\bibitem[Menten et al.(2007)]{men07} {Menten}, K.~M., {Reid}, M.~J., {Forbrich}, J. and {Brunthaler}, A. 2007, A\&A, 474, 515

\bibitem[Michaud, Megessier \& Charland(1981)]{michaud+81} Michaud, G., Megessier, C \& Charland, Y. 1981, A\&A, 103, 244

\bibitem[Morales-Calderón et al.(2011)]{mor11} Morales-Calder\'on, M., Stauffer, J.\,R., Hillenbrand, L.\,A. et al. 2011, ApJ, 733, 50

\bibitem[Morales-Calderón et al.(2012)]{mor12} Morales-Calder\'on, M., Stauffer, J.\,R., Stassun, K.\,G. et al. 2012, ApJ, 753, 149

\bibitem[Muench et al.(2002)]{mue02} Muench, A.\,A., Lada, E.\,A., Lada, C.\,J. \& Alves, J. 2002, ApJ, 573, 366

\bibitem[O'Dell(2001)]{Ode01} O'Dell, C.\,R. 2001, Ann. Rev. A.\&A, 39, 99

\bibitem[O'Dell \& Wen(1994)]{OdW94} O'Dell, C.\,R. \& Wen, Z. 1994, ApJ, 436, 194

\bibitem[Olivares(2012)]{oli12} Olivares, J. 2012, Bachelor's Thesis, UNAM

\bibitem[Olivares et al.(2013)]{oea13} Olivares, J., S\'anchez, L.\,J., Ruelas-Mayorga, A. et al. 2013, AJ, 146, 106

\bibitem[Parenago(1954)]{par54} Parenago, P.~P.\ 1954, Trudy Gosudarstvennogo Astronomicheskogo Instituta, 25, 3

\bibitem[Pecaut \& Mamajek(2013)]{pec13} Pecaut, M.\,J. \& Mamajek, E.\,E. 2013, ApJS, 208, 9

\bibitem[Petr et al.(1998)]{pet98} Petr, M.~G., Coud{\'e} du Foresto, V., Beckwith, S.~V.~W., et al. 1998, ApJ, 500, 825

\bibitem[Preibisch et al.(2005)]{prei05} Preibisch, T., Kim, Y.-C., Favata, F., et al. 2005, ApJS, 160, 401

\bibitem[Prosser et al.(1994)]{pro94} Prosser, C.~F., Stauffer, J.~R., Hartmann, L., et al. 1994, ApJ, 421, 517

\bibitem[Ryabchikova(2014)]{ryab14} Ryabchikova, T., 2014, PSCE.Conf, 220

\bibitem[Ryabchikova et al.(2015)]{ryab15} Ryabchikova, T., Piskunov, N., Kurucz, R.\,L., et al. 2015, PhyS, 90, 054005

\bibitem[Royer et al.(2002)]{roy02} Royer, F., Grenier, S., Baylac, M.-O., Gomez, A.\,E. \& Zorec, J. 2002, A\&A, 393, 897

\bibitem[Searle \& Sargent(1964)]{sea64} Searle, L. \& Sargent, W.~L.~W. 1964, ApJ, 139, 793

\bibitem[Schaller et al.(1992)]{scha92} Schaller, G., Schaerer, D., Meynet, G., et al. 1992, A\&AS, 96, 269

\bibitem[Sicilia-Aguilar et al.(2005)]{sicilia05} Sicilia-Aguilar, A., Hartmann, L.\,W., Szentgorgyi, A.\,H. et al. 2005, AJ, 129, 363

\bibitem[Siess, Dufour, \& Forestini(2000)]{sie00} Siess L., Dufour E., Forestini M., 2000, A\&A, 358, 593

\bibitem[Stelzer et al.(2005)]{ste05} Stelzer, B., Flaccomio, E., Montmerle, T., et al.\ 2005, ApJS, 160, 557

\bibitem[Stickland \& Lloyd(2000)]{stick00} Stickland, D.~J. \& Lloyd, C. 2000, The Observatory, 120, 141

\bibitem[Stift \& Alecian(2012)]{sti12} Stift, M.\,J. \& Alecian, G. 2012, MNRAS, 425, 2715

\bibitem[Tonry \& Davis(1979)]{tonrydavis79} Tonry, J. \& Davis, M. 1979, AJ, 84, 1511

\bibitem[Vitrichenko \& Plachinda(2001)]{vitriplach01} Vitrichenko, E.\,A. \& Plachinda, S.\,I. 2001, Astr. Letters, 27, 581 

\bibitem[Walker(1977)]{wal77} Walker, M.\,F., 1977, Inf. Bull. Var. Stars, 1238

\bibitem[Webb(1859)]{Web59} Webb, T.\,W., 1859, {\it Celestial Objects for Common Telescopes},  London: Longman, Green, Longman, and Roberts

\bibitem[Wolf(1994)]{Wol94} Wolf, G.\,W. 1994, Exp. Astron., 5, 61

\bibitem[Wu et al.(2013)]{Wu13} Wu, Y.\,-L., Close, L.\,M., Males, J.\,R. et al. 2013, ApJ, 774, 45

\bibitem[Wu et al.(2018)]{Wu18} Wu Y.\,-L., Close L.\,M., Kim J.\,S. et al. 2018, ApJ, 854, 144

\bibitem[Zapata, Rodriguez \& Kurtz(2004)]{zap04} Zapata, L.\,A., Rodr{\'\i}guez, L.\,F., Kurtz, S.\,E. et al. 2004, AJ, 127, 2252

\end{thebibliography}
%



\begin{appendix}

\section{Theta 1 Ori F Echelle spectra}
\label{sec:echellespectra}

We present in this Appendix the spectrum resulting from co-adding the four best exposed spectra of $\theta^1$~Ori~F (see Section \ref{sec:spec2}), from 4100\,\AA\ to 6860\,\AA. We omit the sections below and above these limits because they are importantly affected by vignetting and/or low sensitivity; also omitted are those Échelle orders badly contaminated by the strongest nebular emission lines, namely, those between 4740\,\AA\ and 5020\,\AA\ (because of the $H \beta$ and [O\,III] lines), and between 6500\,\AA\ and 6650\,\AA\ (because of the $H \alpha$ and [N\,II] lines). Other obvious nebular lines have been excised from the spectrum. 

\begin{figure*}
\centering

\includegraphics[width=\textwidth]{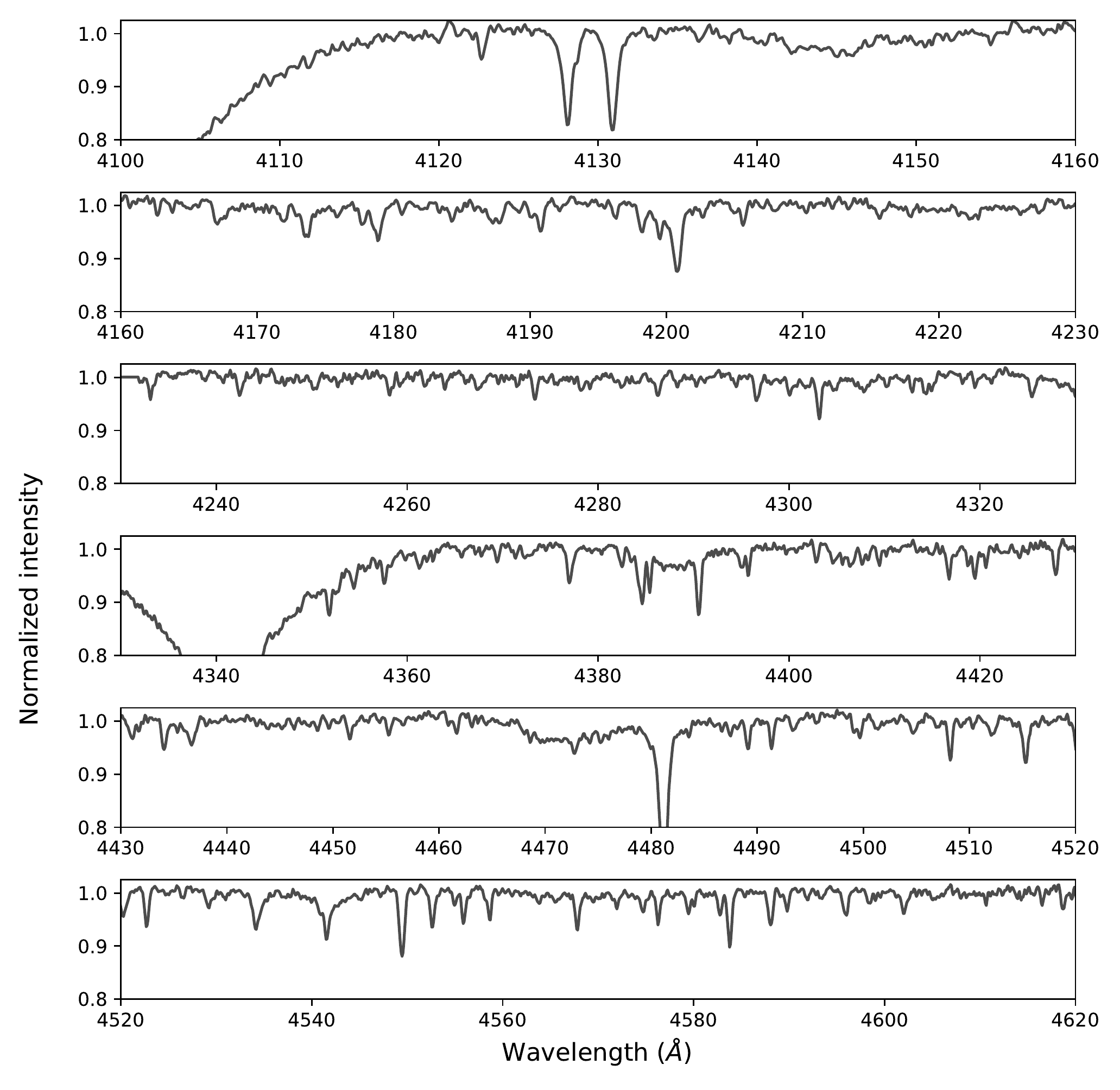}
\caption{$\theta^1$~Ori~F Échelle spectrum from 4100 to 4620\,\AA.  The depth of the strongest metallic line, Mg\,$\rm II$\,4481\,\AA , is 0.68. The wide and shallow features around 4144, 4388 and 4471\,\AA\  are produced by the He\,$\rm I$ lines from the nearby Component C, whose light strongly contaminates the spectrum of the much weaker Component F, especially in the blue Échelle spectral orders.}
\label{fig:echelle1}
\end{figure*}

\begin{figure*}
\centering

\includegraphics[width=\textwidth]{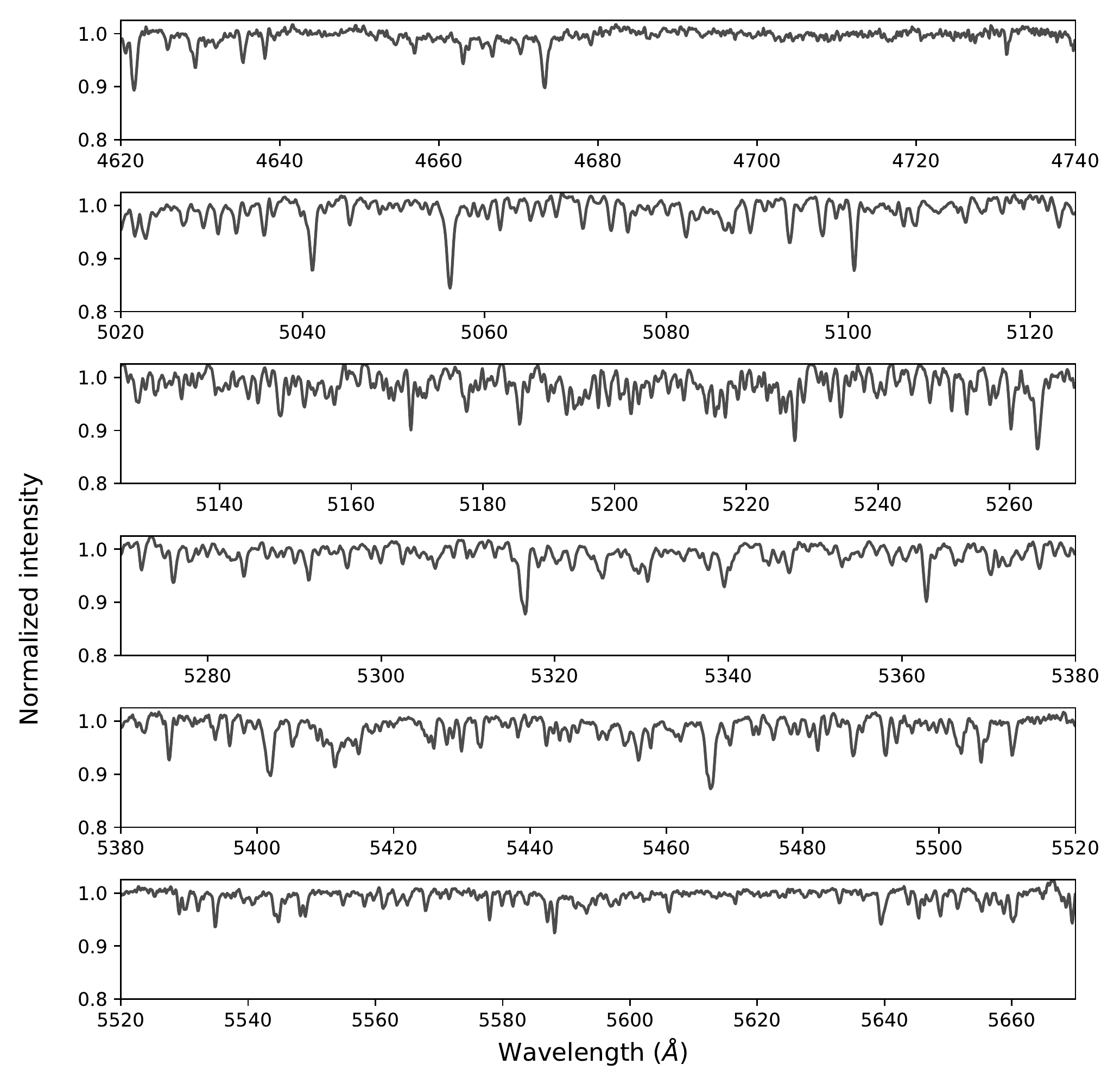}
\caption{$\theta^1$~Ori~F Échelle spectrum, from 4620 to 5665\,\AA.  The region between 4740\,\AA\ and 5020\,\AA is omitted. The relatively wide line at $\lambda$\,5411\,\AA\ is due to He\,$\rm II$ from the contaminating Component C. Noteworthy are the strong Si\,${\rm II}$\,$\lambda\lambda$\,5041, 5056 and 5467\,\AA\ .}
\label{fig:echelle2}

\end{figure*}

\begin{figure*}
\centering

\includegraphics[width=\textwidth]{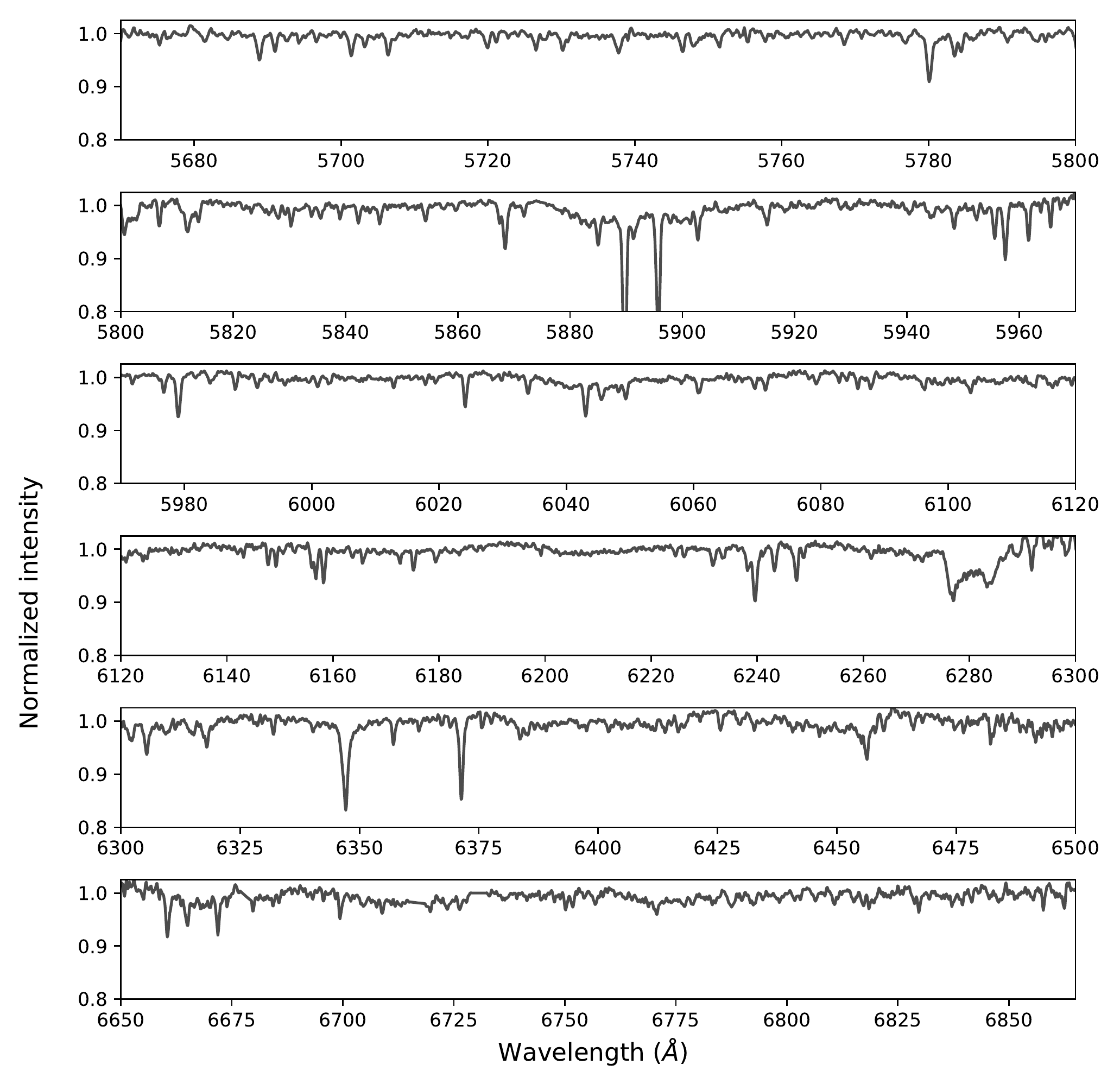}
\caption{$\theta^1$~Ori~F Échelle spectrum from 5670 to 6860\,\AA. The interval between 6500\,\AA\ and 6650\,\AA\ is omitted. Note the abnormally strong P\,${\rm II}$\,$\lambda\lambda$\,6024, 6034 and 6043\,\AA\ lines and the strong Si\,${\rm II}$\,$\lambda\lambda$\,6347, 6371\,\AA\ ones. The very wide and shallow line around the interstellar Na\,${\rm I}$\,$\lambda\lambda$\,5890, 5896\,\AA\ resonant doublet might be due to a reduction artifact (see text).  The deep and wide feature around 6280\,\AA\ is a smeared molecular band of telluric origin.}
\label{fig:echelle3}

\end{figure*}

\end{appendix}


\bsp

\label{lastpage}

\end{document}